\documentclass[reprint,
superscriptaddress,
nofootinbib,
amsmath,amssymb,
aps,
]{revtex4-2}
\usepackage[colorlinks=true,linkcolor=blue,citecolor=blue,urlcolor=
blue]{hyperref}

\usepackage[utf8]{inputenc}
\usepackage{float}
\usepackage{graphicx}
\usepackage{dcolumn}
\usepackage{bm}
\usepackage{amsmath}
\usepackage{tensor}
\usepackage[dvipsnames]{xcolor}
\usepackage{subcaption}
\usepackage{lipsum}
\usepackage{mwe}

\usepackage{graphicx}
\usepackage{subcaption}

\begin{document}
	
	\preprint{APS/123-QED}

    \title{Electromagnetic fields in low-energy heavy-ion collisions with baryon stopping}
    \author{Ankit Kumar Panda}
    \email{ankitkumar.panda@niser.ac.in}
    \author{Partha Bagchi}
    \email{parphy@niser.ac.in}
     \author{Hiranmaya Mishra}
     \email{hiranmaya@niser.ac.in}
    \author{Victor Roy}
    \email{victor@niser.ac.in}

    \affiliation{School of Physical Sciences, National Institute of Science Education and Research, An OCC of Homi Bhabha National Institute, Jatni-752050, India}

	\date{\today}
	
	\begin{abstract}
    We investigate the impact of baryon stopping on the temporal evolution of electromagnetic fields in vacuum at low-energy Au+Au collisions with $\sqrt{s_{NN}} = 4$-$20$ GeV. Baryon stopping is incorporated into the Monte-Carlo Glauber model by employing a parameterized velocity profile of participant nucleons with non-zero deceleration. The presence of these decelerating participants leads to noticeable changes in the centrality and $\sqrt{s_{NN}}$ dependence of electromagnetic fields compared to scenarios with vanishing deceleration. The influence of baryon stopping differs for electric and magnetic fields, also exhibiting variations across their components. We observe slight alteration in the approximate linear dependency of field strengths with $\sqrt{s_{NN}}$ in the presence of deceleration. Additionally, the longitudinal component of the electric field at late times becomes significant in the presence of baryon stopping.

	\end{abstract}

    \maketitle
    \section{Introduction}
  In an off-central relativistic heavy ion collision, intense transient electromagnetic fields are produced predominantly due to the motion of spectators, reaching magnitudes of approximately $\sim 10 m_{\pi}^{2}$ at top RHIC energies~\cite{Skokov:2009qp,Gursoy:2014aka,Voronyuk:2011jd,Deng:2012pc,Kharzeev:2007jp,Roy:2015coa,Alam:2021hje,Zhao:2019crj}. Notably, the peak magnitude of event-averaged values of the electromagnetic field produced in these collisions shows an approximate proportionality to the square root of the center-of-mass energy per nucleon pair, $\sqrt{s_{NN}}$~\cite{Deng:2012pc}. These results, however, were obtained on the assumption of near-perfect transparency of the colliding nucleons, as described by the Glauber model~\cite{Miller:2007ri}. The Glauber model of nucleus-nucleus collisions gives the number of binary collisions and spectators for a given inelastic nucleon-nucleon cross section that depends on $\sqrt{s_{NN}}$ but is assumed to be independent of the number of binary collisions. For high $\sqrt{s_{NN}}$, such as top RHIC and LHC energies, the elastic or diffractive dissociation collisions lead to a minute loss of energy of the colliding nucleon, and hence it is a good approximation to consider that the colliding nucleons move in a straight line with almost constant velocity even after multiple collisions. However, for low $\sqrt{s_{NN}}$ collisions, baryon stopping can be sizeable~\cite{NA49:1998gaz,BRAHMS:2003wwg,BRAHMS:2009wlg,STAR:2017sal}. The baryon stopping must be taken into account while estimating the electromagnetic fields using the Glauber model at low $\sqrt{s_{NN}}$. Particularly, the temporal evolution of the fields post-collisions will be
 affected due to the decelerations of the protons after each binary collision. Previous studies show that, on average, a proton loses half of its pre-collision energy, which is about
one unit of rapidity in each binary collision~\cite{Busza:2018rrf}. The maximum value of the net baryon density at midrapidity is achieved for $\sqrt{s_{NN}} \approx 6 $ GeV~\cite{Cleymans:1992zc,Randrup:2009ch}. Above this collision energy, the midrapidity net baryon density decreases with increasing energy due to the higher transparency of the colliding nuclei. Hence, for higher collision energies, the effect of deceleration in calculating electromagnetic fields is minimal. The decelerated motion of charged protons may additionally give rise to bremsstrahlung radiations that could contribute to direct photon production along with other known mechanisms. 
Experimental measurements on baryon stopping gives an exponential distribution of net baryon density: A exp($-\alpha_B$$\delta $y) where $\delta $y = $Y_{\text{beam}}$- $Y_{\text{CM}}$ is the rapidity loss and the exponent $\alpha_B$ ranges between 0.65-0.67~\cite{BRAHMS:2003wwg,STAR:2008med}. Some theoretical model studies including models based on Color Glass Condensate, could successfully describe these data~\cite{Mohs:2019iee,Li:2018ini,McLerran:2018avb}. However, in the current study, we do not use any of these models; instead, we take a more pedagogical approach and parameterize the decelerated motion of participants in the Monte-Carlo Glauber (MCG)  model to mimic baryon stopping. 

The calculation of precise space-time evolution of electromagnetic fields is vital for studying anomaly-induced parity-violating phenomena in QCD, such as the chiral magnetic effect (CME), where a charge/chiral current current 
 $\boldsymbol{j}/ \boldsymbol{j_5}=\frac{e^2}{2 \pi^2} (\mu_5/\mu) \boldsymbol{B}$
 (here $\mu_5$ and $\mu$ are chiral and charge chemical potential respectively)  is induced along magnetic fields in a system with chiral fermions. This phenomenon has attracted significant attention since its inception~\cite{Kharzeev:2004ey,Fukushima:2008xe,Kharzeev:1998kz,Huang:2015oca,ALICE:2023weh,STAR:2021mii,STAR:2022ahj,Hattori:2022hyo}. As an experimental signature the charge-dependent two- and three-particle correlations in Xe-Xe collisions has been proposed as a signature to the presence of CME. However, one needs to separate the signal of CME from the possible background and compare the experimental data with theoretical model studies that take into account the correct spatiotemporal evolution of electromagnetic fields. The lifetime and intensity of the electromagnetic fields in heavy-ion collisions and hence the CME signal depends on the center of mass energy of the collision and the chiral chemical potential of the hot and dense nuclear matter, among other deciding factors. The Beam Energy Scan (BES) program at RHIC and the future Facility for Antiproton and Ion Research (FAIR) at GSI aim to explore the properties of the QCD matter at several low center of mass energy collisions. These experiments will also provide a unique opportunity to study whether the CME and related phenomena could still be observed in lower-energy heavy-ion collisions. 
 
As mentioned earlier, we need a dynamical model that captures the space-time evolution of electromagnetic fields with the initial fields calculated from an initial model such as MC Galuber and solve the Maxwell and fluid equations self-consistently to represent the space-time evolution faithfully. As a first step towards that direction, we study the temporal evolution of electromagnetic fields in vacuum for lower $\sqrt{s_{NN}}$ collisions with baryon stopping. We mimic the effect of baryon stopping via a parameterized velocity profile (with non-zero deceleration) of nucleons undergoing binary collisions. 
  
  Earlier studies related to the evolution of electromagnetic field in conducting medium~\cite{Kharzeev:2007jp,PhysRevC.107.034901} at rest and in an expanding charged conducting fluid~\cite{Dash:2023kvr} show that the fields are sustained for a more extended period of time compared to the vacuum evolution. Further, we expect a longer-lasting electromagnetic field in low-energy collisions, albeit with a smaller peak value, which is an interesting unexplored territory that we wish to explore. Specifically, the effect of fields on anisotropic flow in low-energy heavy-ion collisions is worthwhile to study as their high-energy counterpart shows some interesting dependencies on flow coefficients and transport properties~\cite{STAR:2023jdd,Tuchin:2014hza,Das:2017qfi,Denicol:2018rbw,Denicol:2019iyh,Dubla:2020bdz,Das:2022lqh,Panda:2023akn,Inghirami:2019mkc,Panda:2020zhr,Panda:2021pvq,Ambrus:2022vif,Dash:2022xkz,Nakamura:2022wqr,Nakamura:2022idq,Nakamura:2022ssn,Kushwah:2024zgd,Singh:2024leo,Singh:2023pwf}. These studies will give us unique access to measure the transport coefficients of hot hadronic gas.

This paper is organized as follows: In the first section, in Sec.\eqref{sec:formulations}, we discuss the theoretical formulation and the assumptions made for introducing the stopping of baryons after the collisions. Then, in Sec.\eqref{montecarlo}, we briefly discuss the Monte-Carlo Glauber model. Next, in Sec.\eqref{results}, we present the results. Finally, we conclude in Sec.\eqref{conclusion}. Throughout the paper, we use natural units, $\hbar=c=k_{B}=\epsilon_{0}=\mu_{0}=1$.

     \section{Formulations}
     \label{sec:formulations}
As mentioned in the introduction, we consider the deceleration of charged participants after they undergo binary collision. The electromagnetic fields for a point particle with charge $Ze$ moving with velocity $\bm{{\bm{{{\beta}}}}}=\frac{d \bf{r}}{dt}$ and a proper acceleration $\bm{\dot{{\bm{\beta}}}}= \frac{d \bf{{\bm{\beta}}}}{dt}$ can be calculated at position $ {\bf{r}}_{\text{obs}}$ at time $t_{\text{obs}}$ from the well known formula~\cite{lecturenotes}.
\begin{widetext}
  \begin{equation}{\label{eq:fields}}
    \begin{aligned}
        e{\bf{B}}({\bf{r}}_{\text{obs}},t_{\text{obs}}) &= -\mathcal{C}  Z \alpha_{EM} \left[ \frac{\hat{\bf{R}} \times {\bm{\beta}(t')}}{\gamma^2 k^3 R^2} + \frac{(\hat{{\bf R}} \cdot \dot{\bm{\beta}}(t')) (\hat{\mathbf{R}} \times {\bm\beta}(t')) + k \hat{{\bf R}} \times \dot{{\bm \beta}}(t')}{k^3 R}\right]_{t'} ,\\
        e{\bf{E}}({\bf{r}}_{\text{obs}},t_{\text{obs}}) &= \mathcal{C} Z \alpha_{EM}  \left[ \frac{\hat{{\bf{R}}}-{\bm\beta}(t')}{\gamma^2 k^3 R^2} + \frac{ \hat{\bm{R}} \times [(\hat{\bf{R}}-{\bm{\beta}}(t'))\times \dot{{\bm{\beta}}}(t')]}{k^3 R } \right]_{t'},
    \end{aligned}
\end{equation}
\end{widetext}
where $e$ is electronic charge, the fields are all in units of $m_{\pi}^2$, and $R$ is in fm.
Here $\mathcal{C}=fm^{-2}/m_{\pi}^{2} \sim 2$ is a numerical factor, $\alpha_{EM}= \frac{1}{137}$ is the fine structure constant, and $Z$ is the atomic number of each nucleus (we consider symmetric collisions). The right-hand side of the above expressions is evaluated at retarded time $t^{\prime}$. The relation between $t'$ and $t_{\text{obs}}$ is given by
 \begin{equation}{\label{eq:retardedt}}
    t' + \sqrt{(x_{\text{obs}}-x^{\prime})^2 + (y_{\text{obs}}-y^{\prime})^2 + (z_{\text{obs}}- z^{\prime}(t'))^2} = t_{\text{obs}}.
  \end{equation}
  The relative position $\bf{R}(t^{\prime}) = \bf{r}_{\text{obs}}-\bf{r'}(t')$, and the unit vector along it is defined as $\hat{\bf{R}} = \frac{\vec{R}}{R}$, the factor $k = 1- \hat{\bf{R}} \cdot \bm{\beta}(t')$, and $\gamma= \frac{1}{\sqrt{1-\beta^2}}$ is the Lorentz factor.


To calculate the electromagnetic fields, we consider nucleons moving in a straight line with a constant velocity $\beta_{s_{NN}} = \left(1.0-\frac{4 m^2}{s_{NN}}\right)^{1/2}$, $m$ is the mass of the proton. Depending on whether they are participants or spectators, we decelerate or let them continue moving with constant
initial velocity for a given $\sqrt{s_{NN}}$. Below, we discuss a step-by-step process for evaluating the total electromagnetic field using an MC Glauber model and parameterized form for deceleration. 
\begin{itemize}
 \item We sample the positions of individual nucleons from the nuclear density distribution of the Wood-Saxon type (the details of which are discussed later). This will give us the initial positions $x^{\prime}_0(t')$, $y^{\prime}_0(t')$ and $z^{\prime}_0(t')$ of the nucleons inside the right and left moving nucleus. We assume an Eikonal approximation where, even after binary collisions, individual nucleons will continue moving along the beam's direction.
  So far, the calculation of the fields is done at some retarded time $t^{\prime}$, and then we obtain the corresponding fields at $t_{\text{obs}}$ using Eq.\eqref{eq:retardedt}. Here, as we will be working with the collision of individual nucleons tracking its trajectory, it is more convenient to work in proper time $\tau$ where $\tau= \sqrt{t^2-\mathbf{r}^2}$. Here, we choose $\tau=0$ at $\mathbf{r}=0$, corresponding to $t=0$ when the centers of the two nuclei overlap. In this scenario, individual nucleon-nucleon collisions can occur for $\tau <0$ or $\tau>0$, assuming that just after the collision, the nucleons that take part in the collision process would decelerate.
    \item If the proton is a participant, we apply a deceleration with the following parametrized form for the velocity profile as
  \begin{eqnarray}{\label{eq:partvel}}
  \beta(\tau) = \mathcal{A} \left[1 - \tanh\left( \frac{\tau - \tau_{h}}{\Delta\tau} \right)\right],
\end{eqnarray}
where, $\mathcal{A} = \frac{\beta_{s_{NN}}}{2}$, $\Delta\tau$ is a parameter used to control the time interval for the deceleration, $\tau_h$ is a parameter to control the time scale of the deceleration such that at $\tau=\tau_h$ the initial velocity $\beta_{s_{NN}}$ is reduced to half of its original value. Further we define a starting time $\tau_s$ for individual nucleon-nucleon collision as the time when $\beta_{s_{NN}}/\beta(\tau_s)\sim0.98$. As per our definition, $\tau_s$ could be positive or negative depending on the location of the participants inside the colliding nucleus. The choice of the starting time of collision, where velocity reduces to $98\%$ of the initial velocity, is reasonable. Firstly, it helps us to handle the discontinuity in velocities at the onset of deceleration. Moreover, from a physical perspective, this reduction can be interpreted as arising from Coulombic repulsion or due to the composite structure of the nucleons. We further note that the definition $\tau_s$ is arbitrary. Still, it is a reasonable choice because we assume the nucleons in the MC Glauber model are hard spheres with radius given by the inelastic nucleon-nucleon cross section as given later in Eq.\eqref{eq:rc}. A collision happens when they touch each other.  

The starting time for individual nucleon-nucleon collisions in a given event is calculated based on their relative distances $\Delta z = z_{target} - z_{projectile}$ at $\tau=0$, and considering straight line trajectories with velocity $\beta_{s_{NN}}$.
  \begin{figure}[H]
    \centering
    \includegraphics[width=1\linewidth]{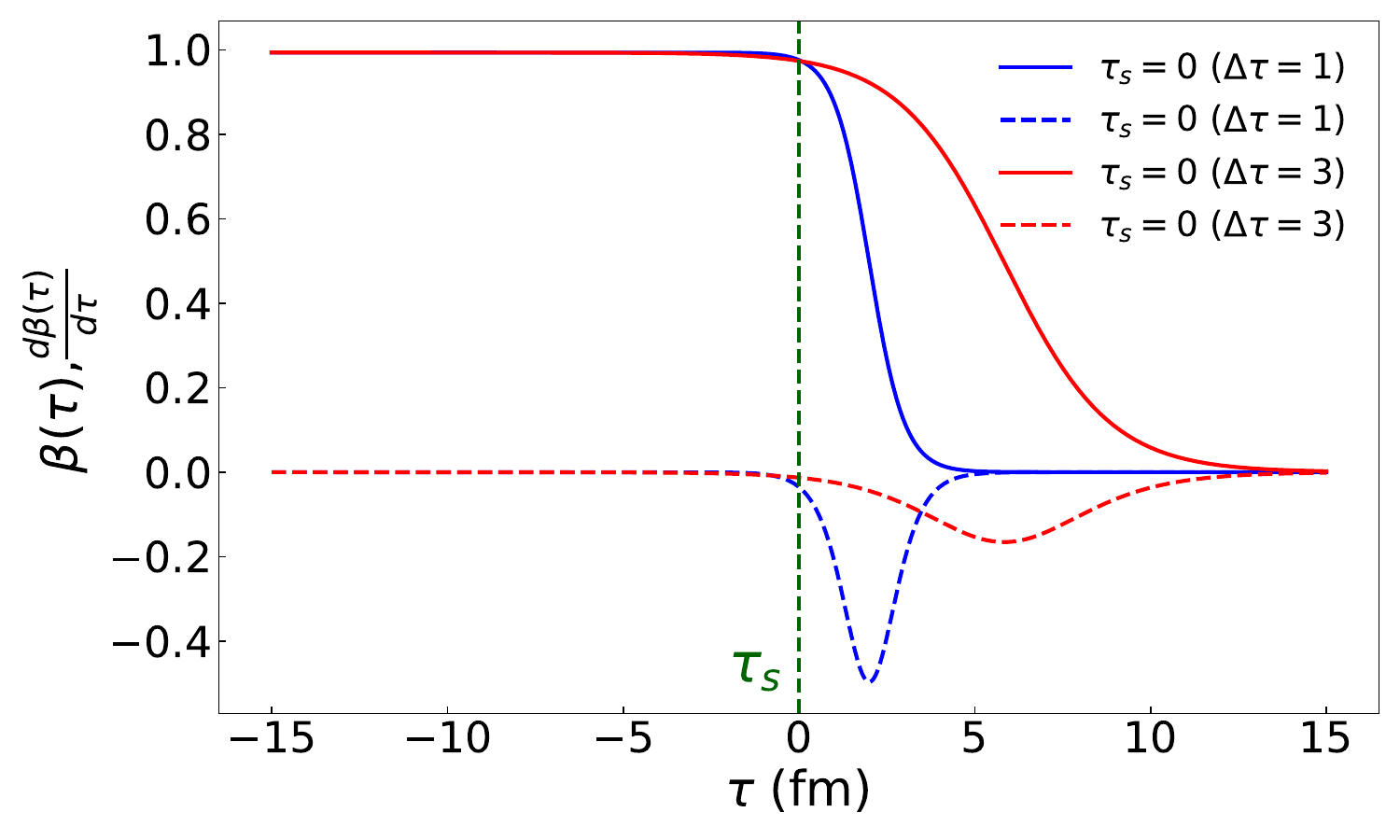}
  \caption{(Color online) Parametrization of velocity $\beta$ ($\tau$) (solid lines) and the corresponding $\frac{d\beta}{d\tau}$ (dashed lines) as a function of $\tau$ for participants for $\Delta \tau$ = 1 fm (blue curves) and 3 fm (red curves). }\label{fig:velpar}
\end{figure}
 In Fig.\eqref{fig:velpar} we display the velocity profile $\beta(\tau)$ (solid lines) and the proper acceleration $\frac{d\beta}{d\tau}$ vs $\tau$ (dashed lines) for $\Delta \tau$ =1 fm (blue lines) and 3 fm (red lines) respectively. 
    \item In Eq.\eqref{eq:fields}, EM fields are evaluated at retarded time; hence we need to find the retarded time and corresponding positions 
    of the nucleons from $\beta{(\tau)}$ using the relations $$\frac{d t'(\tau)}{d \tau}= \frac{1}{\sqrt{1-\beta^2 (\tau)}}, \frac{d z' (\tau)}{d \tau}= \frac{\beta (\tau)}{\sqrt{1-\beta^2 (\tau)}}. $$
    \item Finally, we obtained the electromagnetic fields at observation point ${\bf{r}}_{\rm obs}\equiv (x,y,z)$ at present $t_{\text{obs}}$ from Eq.\eqref{eq:fields}. Calculating the electromagnetic field in an event-by-event case may result in some nucleons being very close to the point of observation, making $|\mathbf{R}| \approx 0$ leading to divergence for the fields. In practical calculations, different regularisation schemes have been used, and consistent results are obtained after taking the event average~\cite{Deng:2012pc,Deng:2014uja,Skokov:2009qp,Voronyuk:2011jd}.  To address this issue, here we introduce a cutoff at $|\mathbf{R}| \approx$ 1 fm. This implies that the fields associated with nucleons within a distance of $|\mathbf{R}| = 1$ fm are discarded.

  \end{itemize}
  
  In this work, we calculate EM fields from an ensemble of a thousand events for a given collision centrality.   
    

\section{Monte-Carlo Glauber}\label{montecarlo}

\begin{figure}[H]
    \centering
    \includegraphics[width=1.0\linewidth]{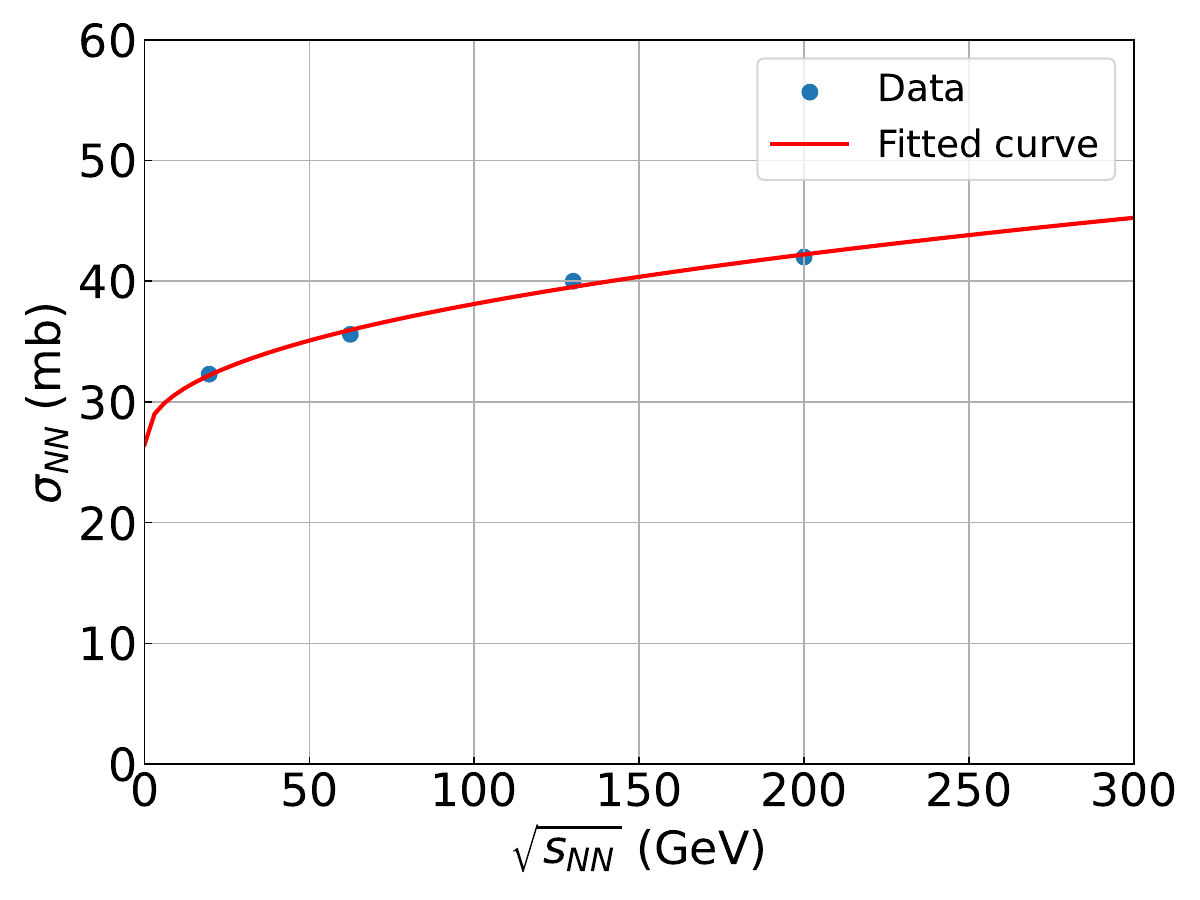}
    \caption{(Color online) $\sigma_{NN}$ (nucleon-nucleon inelastic cross-section) vs $\sqrt{s_{NN}}$ (circles) taken from~\cite{Miller:2007ri}, the red line is fit with a three parameter function Eq.\eqref{eq:fit}.}
    \label{fig:sigmaNN} 
   \centering
  \includegraphics[width=1.0\linewidth]{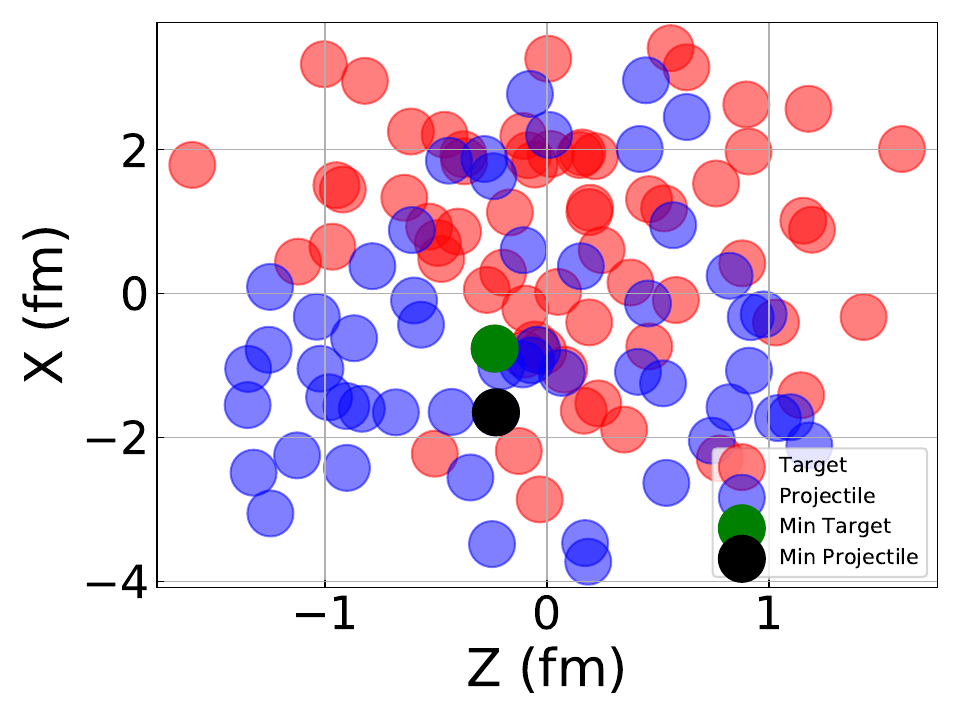}
    \caption{(Color online) Snapshot of the participants  in a given Au+Au collision event at $\tau$ $=$ 0 for $\sqrt{s_{NN}}$= 8 GeV and $b$=8 fm. }
    {\label{fig:targandproj}}
\end{figure}

As mentioned earlier, we use the MC Glauber model~\cite{Miller:2007ri} to calculate the nucleon distributions, participants, spectators, and number of binary
collisions for a given $\sqrt{s_{NN}}$ and impact parameter, here we briefly discuss the essential features and parameters used in our study. 
We sample the nucleon positions inside a given nucleus from the corresponding Wood-Saxon density distribution (assuming spherical symmetry):
\begin{eqnarray}
  \rho(r) &=& \frac{\rho_0}{1+ e^{\frac{r-R}{a}}}.
\end{eqnarray}
Where $\rho_0$ is the nucleon density in the center of the nucleus, R is the radius of the nucleus, $a$ is the skin depth, and $r$ is the radial distance from the center of the nucleus. For Au$^{197}$ we use: $\rho_0 = 0.16$ fm$^{-3}$, $R= 6.34$ fm, a = 0.54 fm. We calculate participant for a given nucleon-nucleon inelastic cross section $\sigma_{NN}$ by considering individual nucleons as a hard sphere; a collision takes place if the inter-nucleon transverse distance $r_{\bot}= \sqrt{(x_p-x_T)^2 + (y_p- y_T)^2} \leq r_c$, where 
\begin{eqnarray}
\label{eq:rc}
  r_{c} &=& \sqrt{\frac{\sigma_{NN}}{\pi}}.
\end{eqnarray}
The ($x_p$, $y_p$) and ($x_T$, $y_T$) mentioned above are the transverse positions of projectile and target nucleons respectively.
The experimentally measured values of $\sigma_{NN}$ are available for selected energies, we fit the experimentally 
measured $\sigma_{NN}$ vs $\sqrt{{s}_{NN}}$ with the following three parameters form,
\begin{eqnarray}
\label{eq:fit}
  \sigma_{NN}^{f}&=& A \left(\sqrt{{s}_{NN}}\right)^B + C.
\end{eqnarray}
Here, from the fit we obtain the values $A = 7.63$, $B = 0.22$, $C = 17.36$ with $\chi^2 \approx 0.05$. 
The parametric fit and experimental data points (circles) are shown in Fig.\eqref{fig:sigmaNN}.
It is worthwhile noting that throughout the rest of the paper, we assume that the trajectory of the participants is governed by the Eq.\eqref{eq:partvel} and each of the participants will eventually lose much of its energy within a time interval $\Delta \tau$. However, the baryon usually loses about one unit of rapidity in each collision, and it will take multiple collisions before they lose a substantial amount of the initial energy \cite{PhysRevLett.52.1393, PhysRevD.39.2606}. Such a realistic scenario of energy/rapidity loss is beyond the scope of the present work and can be implemented in a future study. 

\section{Results and Discussions}\label{results}
From now on, we denote `electronic charge  times the fields' simply as `Fields' in our plots.
We primarily focus on event-averaged values of the electric ${\bf{E}}\equiv(E_x,E_y,E_z)\equiv E_i$ and magnetic ${\bf{B}}\equiv (B_x, B_y, B_z) \equiv B_i $  field components to investigate the effect of baryon stopping on the electromagnetic fields unless stated otherwise. For the event-averaged case, we take ensemble of a thousand events, and to focus on the event-by-event contribution of each field component, we consider their absolute values so that the random phase cancellation during the averaging could be avoided.
\begin{figure}[h!]
    \centering
    \includegraphics[width=0.8\linewidth]{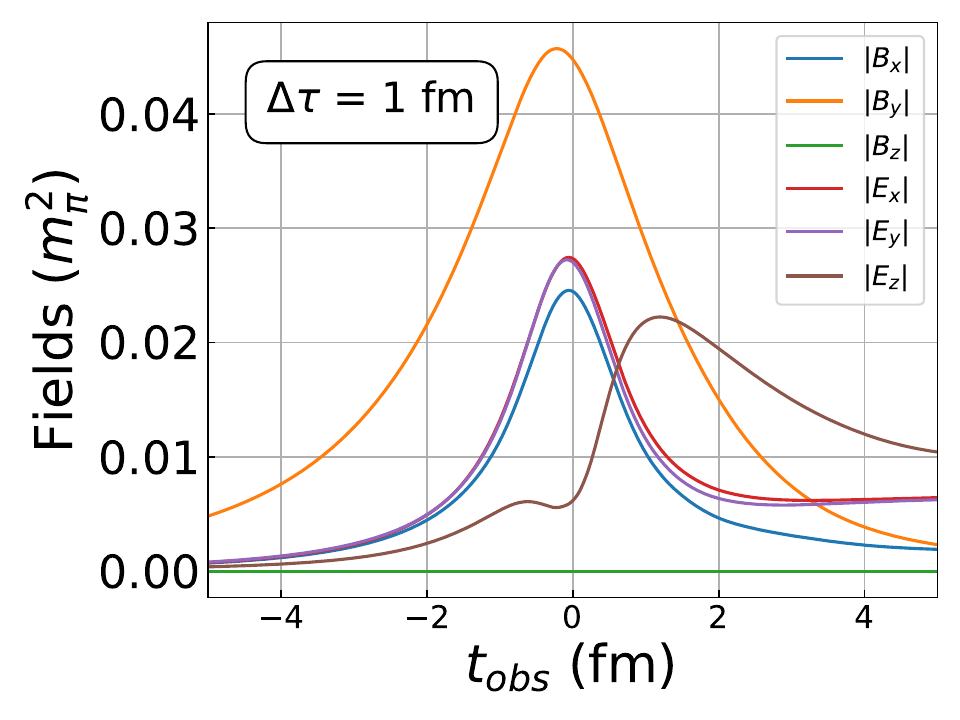}
    \centering 
    \includegraphics[width=0.8\linewidth]{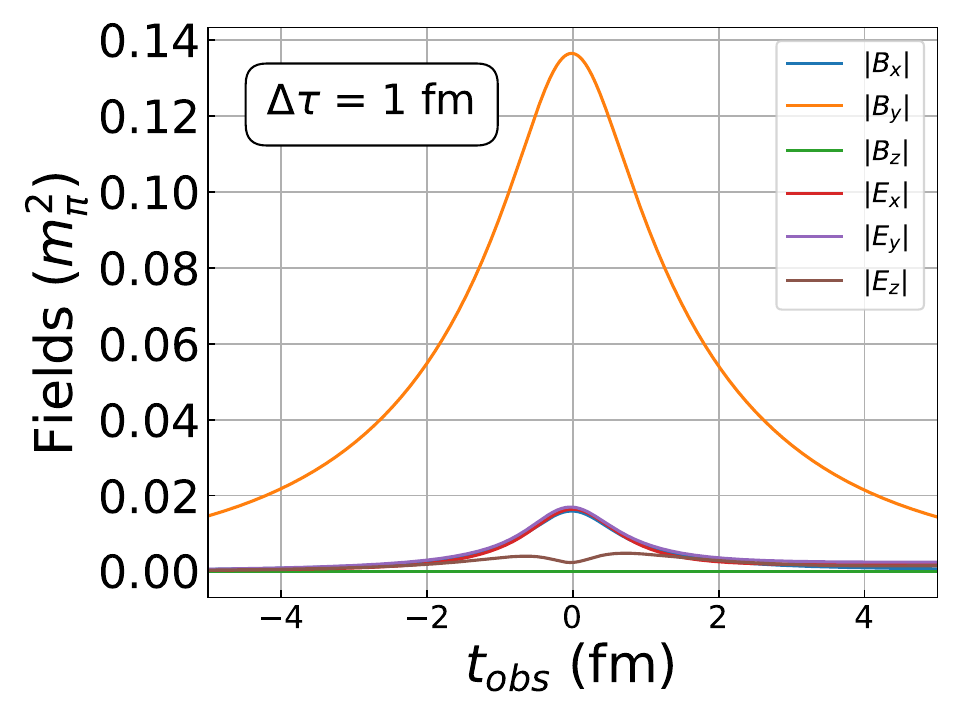}
  \caption{(Color online) Evolution of ${\bf E}, {\bf B}$  at $\textbf{r}_{\text{obs}} = (0, 0, 0)$ over time for $b = 3$ (\text{top}) and $12$ fm (\text{bottom}) at $\sqrt{s_{NN}} = 4$ GeV, with $\Delta \tau = 1$ fm.
}\label{fig:temporalevolution}
\end{figure}

Since the number of participants/collisions decreases monotonically with the impact parameter/centrality of the collision, we expect the effects of
baryon stopping with deceleration on the fields is minimal for peripheral collisions. 
Fig.\eqref{fig:temporalevolution} depicts the temporal evolution of $|E_i|$ and $|B_i|$ at the center of the collision zone ${\bf r}_{\text{obs}} =(0,0,0)$, for $\Delta \tau = 1 \, \text{fm}$ at $\sqrt{s_{NN}} = 4 \, \text{GeV}$ for two distinct impact parameters: 3 fm and 12 fm, respectively. Upon comparison of both plots, a notable distinction emerges, particularly evident in the more central collision (at $b = 3 \, \text{fm}$), showcasing the effect of deceleration being dominant at the most central collision which can be attributed to the increased number of participants. 


\begin{figure}[H]
    \centering
    \includegraphics[width=0.8\linewidth]{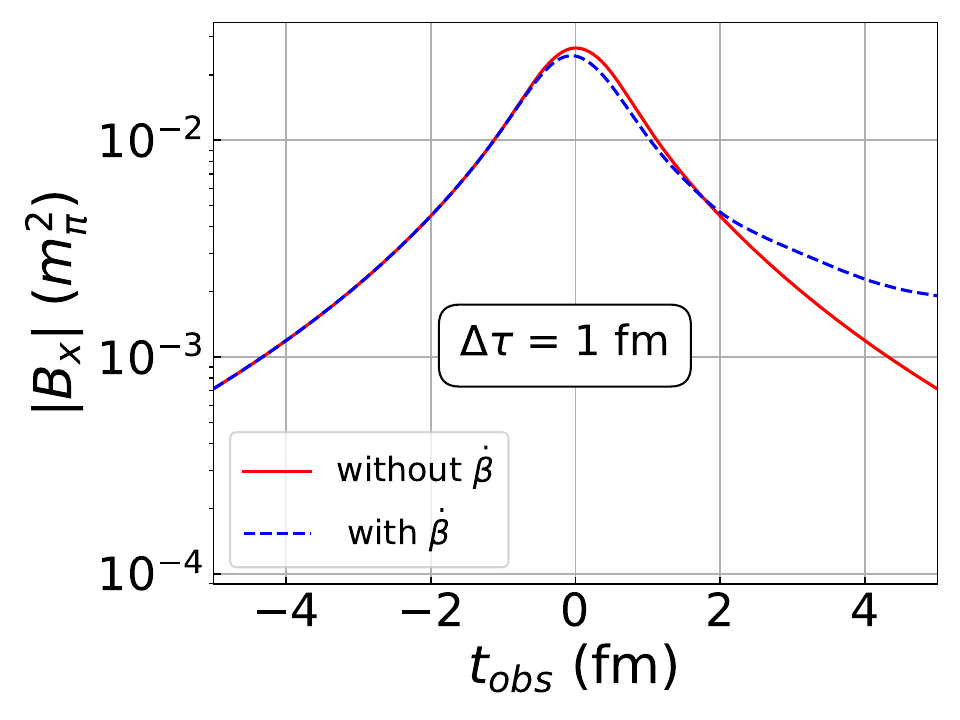}
    \centering
    \includegraphics[width=0.8\linewidth]{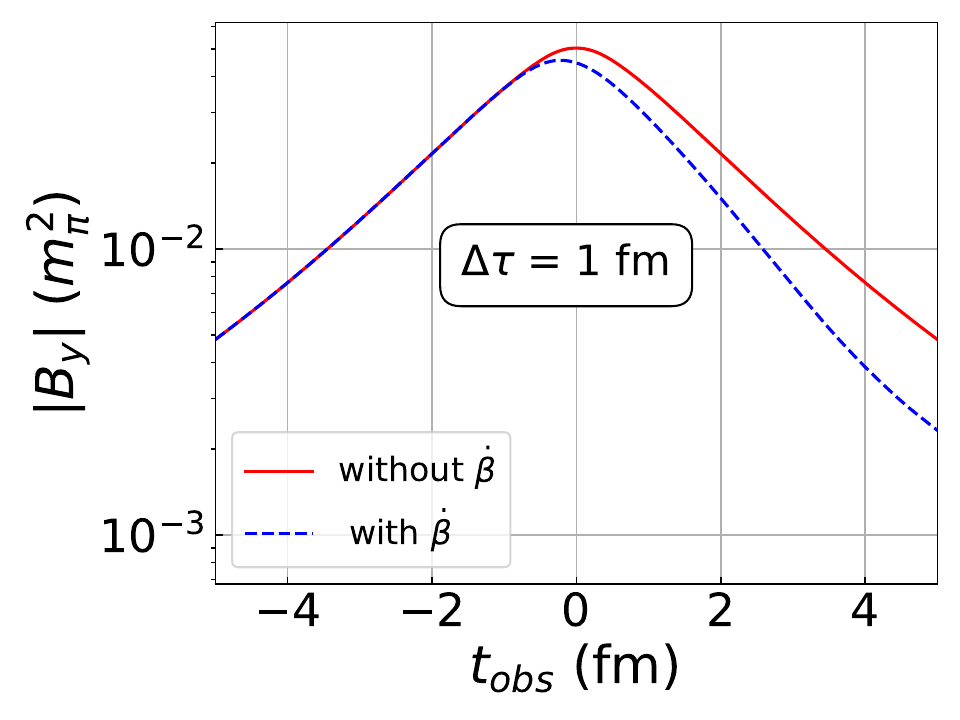}
    \centering
    \includegraphics[width=0.8\linewidth]{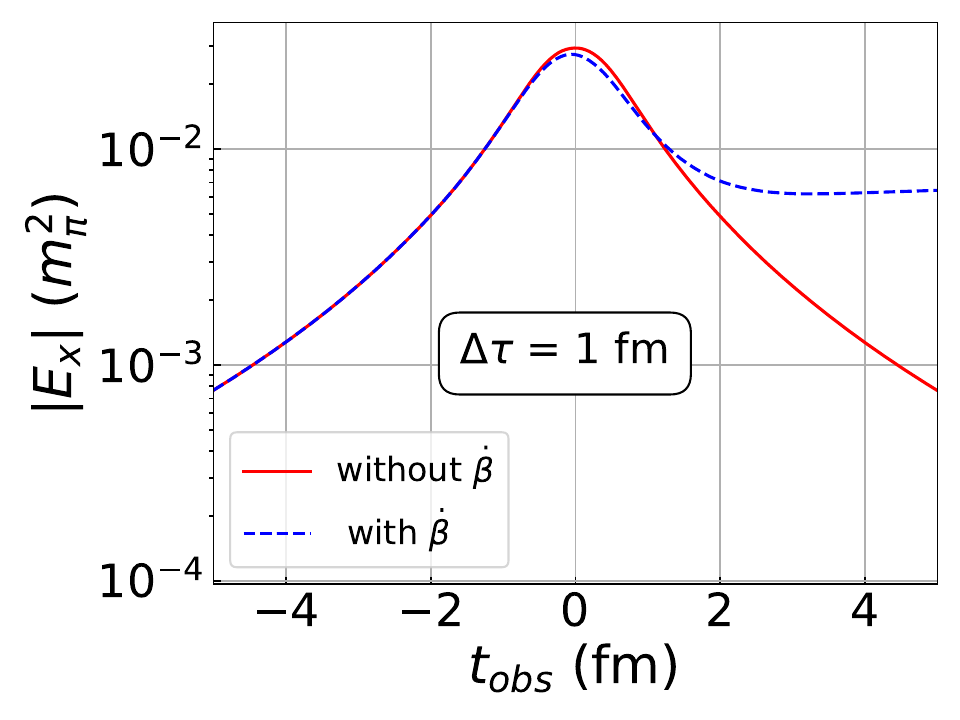}
    \centering
    \includegraphics[width=0.8\linewidth]{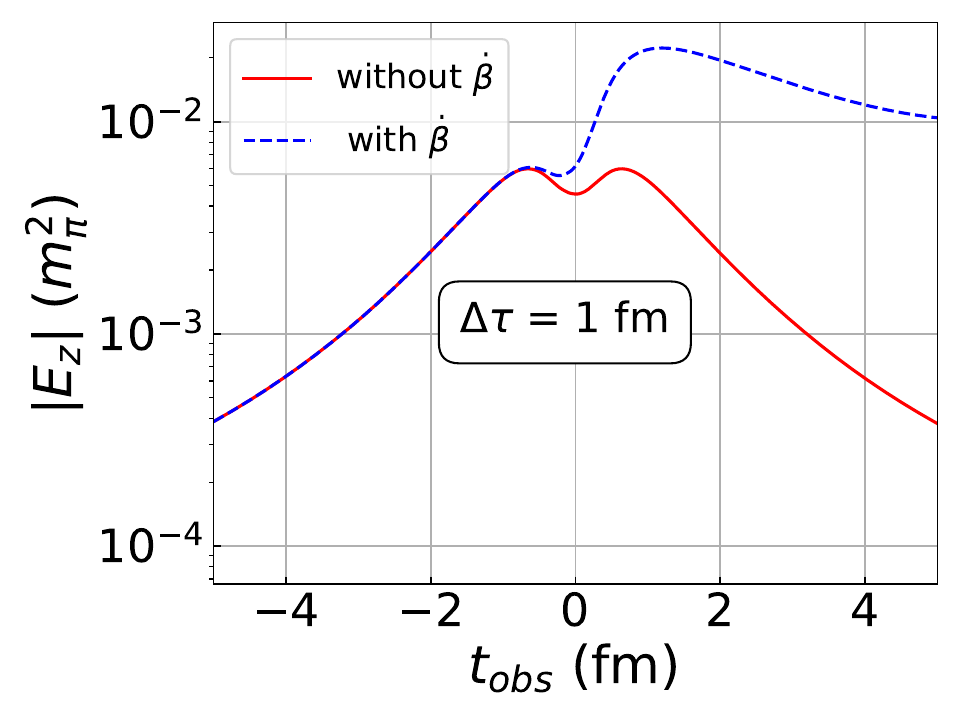}
  \caption{(Color online) Comparison of $|B_x|, |B_y|$ (top two panels), and  $|E_x|, |E_z|$ (bottom two panels) with (blue dashed lines) and without deceleration (red solid lines) at $\sqrt{s_{NN}} = 4$ GeV with $b$ = 3 fm at $\textbf{r}_{\text{obs}}= (0,0,0)$.}\label{fig:compwithwithout}
\end{figure}
In the presence of deceleration the magnitude of $|B_x|$, $|E_x|$, $|E_y|$, and $|E_z|$ are enhanced in the presence of deceleration after $\tau \sim$ 2 fm. Furthermore, electric fields seem to asymptotically reach a constant value for $t_{\text{obs}} \gtrapprox 4 \, \text{fm}$.
 This late-time behavior arises from the dominance of Coulombic fields as deceleration drives participant velocities towards non-relativistic limits, and they may eventually come to rest. Additionally, a slight shift in the peak values of $|B_x|$, $|B_y|$, $|E_x|$, and $|E_y|$ is noticeable in the \text{top plot} compared to the \text{bottom}, primarily attributed to the velocity profile of participants which start decelerating slightly before the collision, i.e., $\tau < 0 $.
To see the effect of baryon stopping, we show the comparison of the temporal evolution of 
the electromagnetic fields with and without baryon stopping at $\sqrt{s_{NN}} = 4$ GeV 
for $b = 3$ fm in Fig.\eqref{fig:compwithwithout}. The solid red lines correspond to no stopping, and the blue dashed lines 
correspond to baryon stopping. It is evident that $|E_x|$ and $|E_z|$ are significantly higher for the stopping scenario at late times (after 3-4 fm).
 Whereas $|B_y|$ does not show this type of asymptotic behavior at late times, as seen from the second plot from the top in Fig.\eqref{fig:compwithwithout}. This is understood as charges at rest won't give rise to any magnetic fields, unlike the Coulombic contribution to the electric fields. The other components of the magnetic fields show similar time dependence.

\begin{figure}[H]
    \centering
    \includegraphics[width=0.85\linewidth]{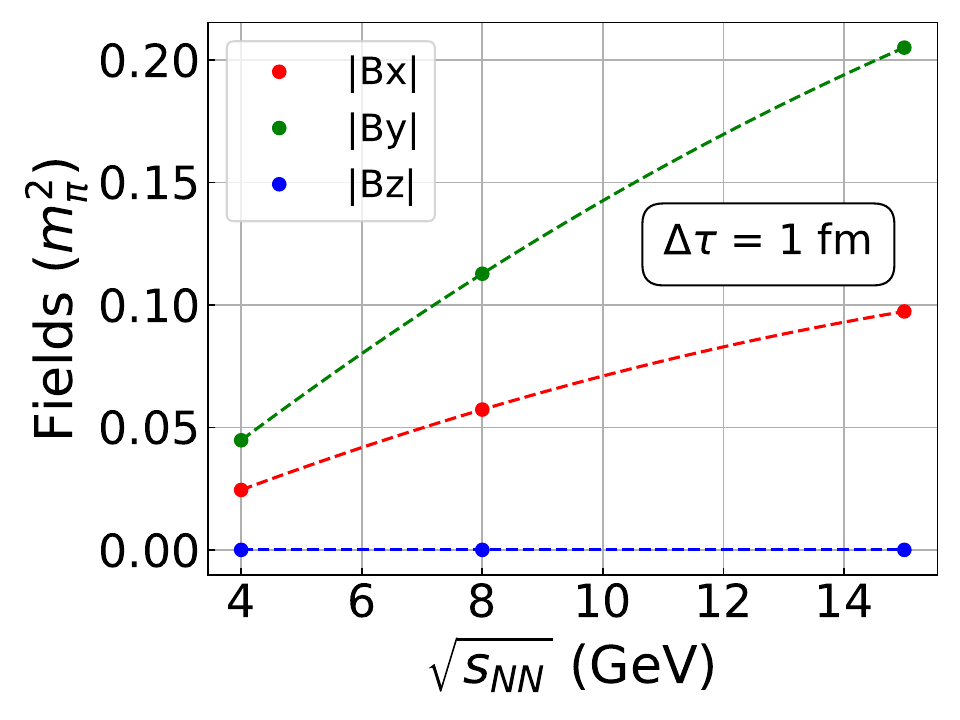}
    \centering
    \includegraphics[width=0.85\linewidth]{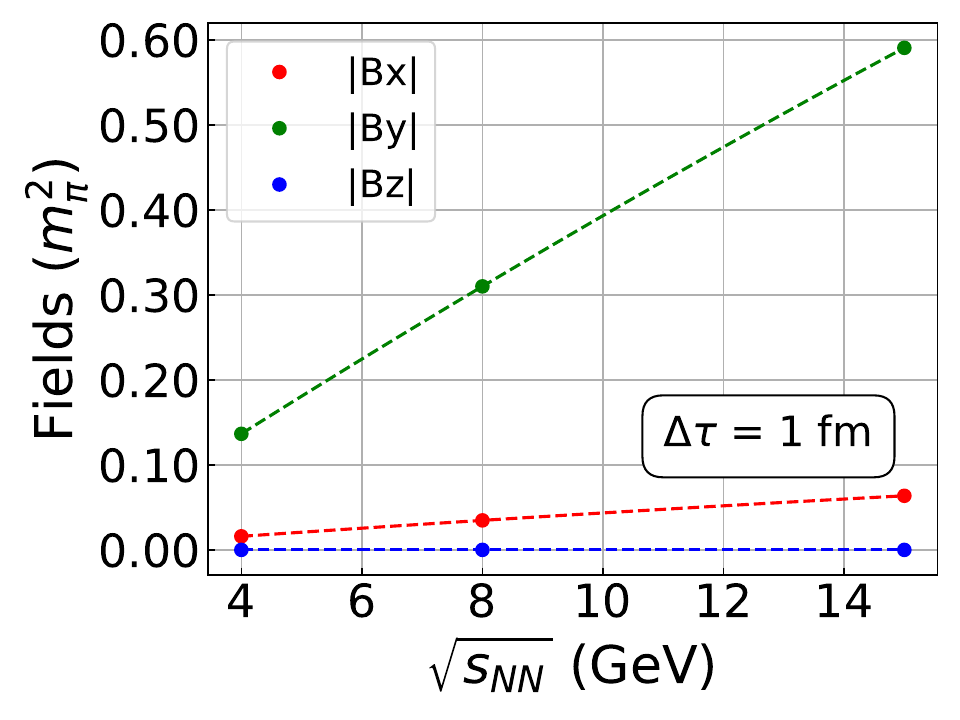}
  \caption{(Color online) Variation of magnetic fields with $\sqrt{s_{NN}}$ for $b$= 3 (top panel), 12 fm (bottom panel) at ${\bf r}_{\text{obs}}$ = (0,0,0).}\label{fig:rootsvariation}
\end{figure}
\begin{figure}[H]
    \centering
    \includegraphics[width=0.85\linewidth]{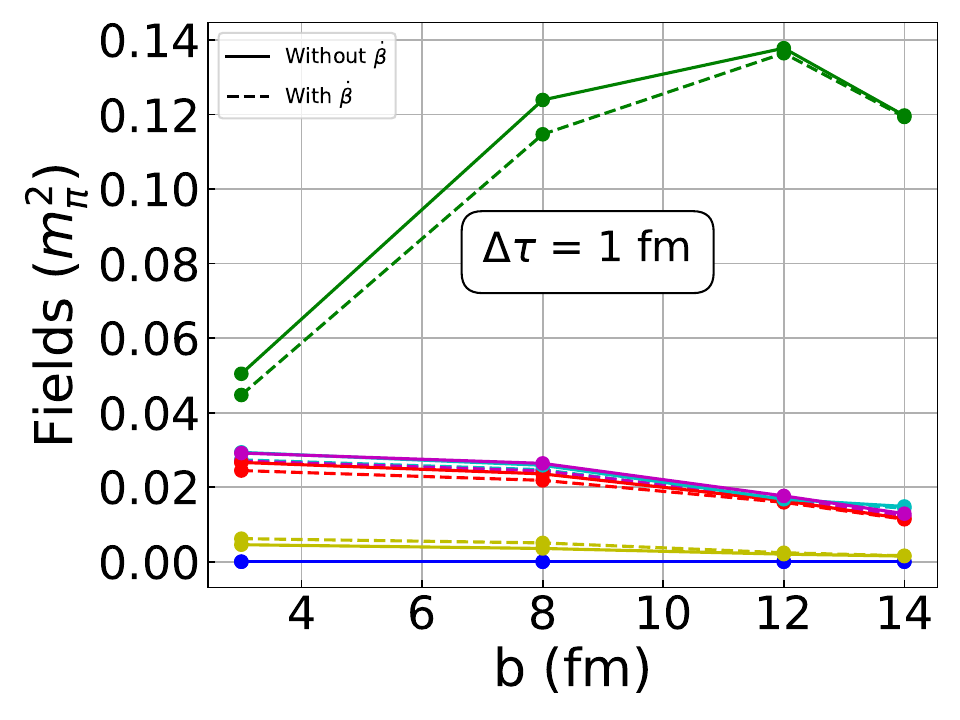}
  \caption{(Color online) Variation of $B_x$ (red), $B_y$ (green) ,$B_z$ (blue),  $E_x$ (cyan), $E_y$ (magenta), $E_z$ (yellow)  with $b$ for $\sqrt{s_{NN}}$ = 4 at $(t,{\bf {r}_{\text{obs}}})$ = $(0, {\bf {0}})$ with (dashed lines) and without deceleration (solid lines). }\label{fig:variationwithb}
\end{figure}
\begin{figure}[H]
    \centering
    \includegraphics[width=0.8\linewidth]{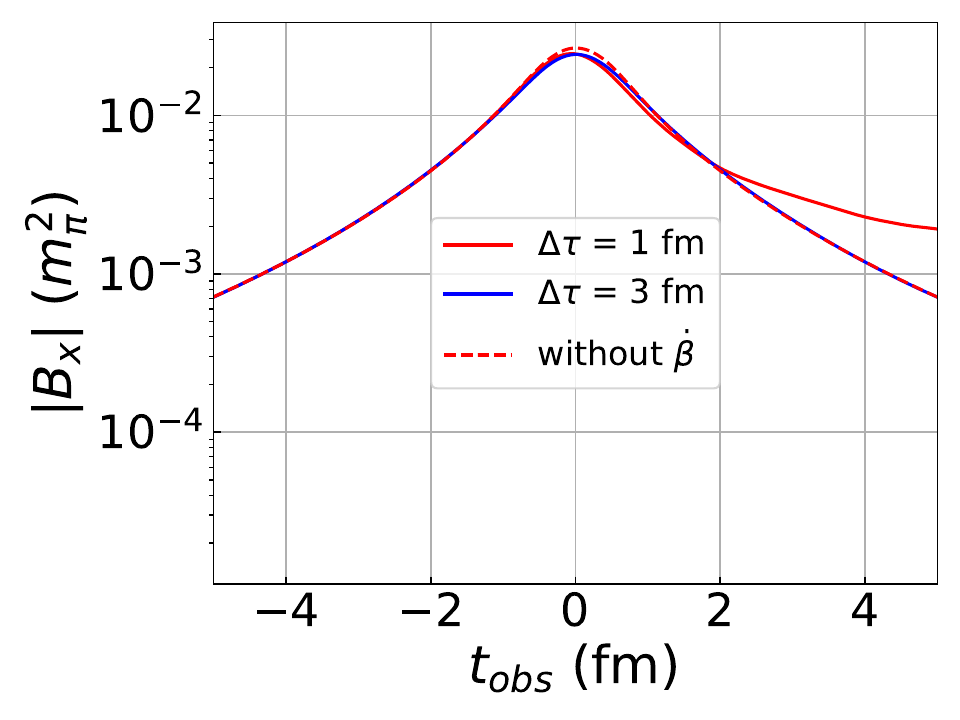}
    \centering
    \includegraphics[width=0.8\linewidth]{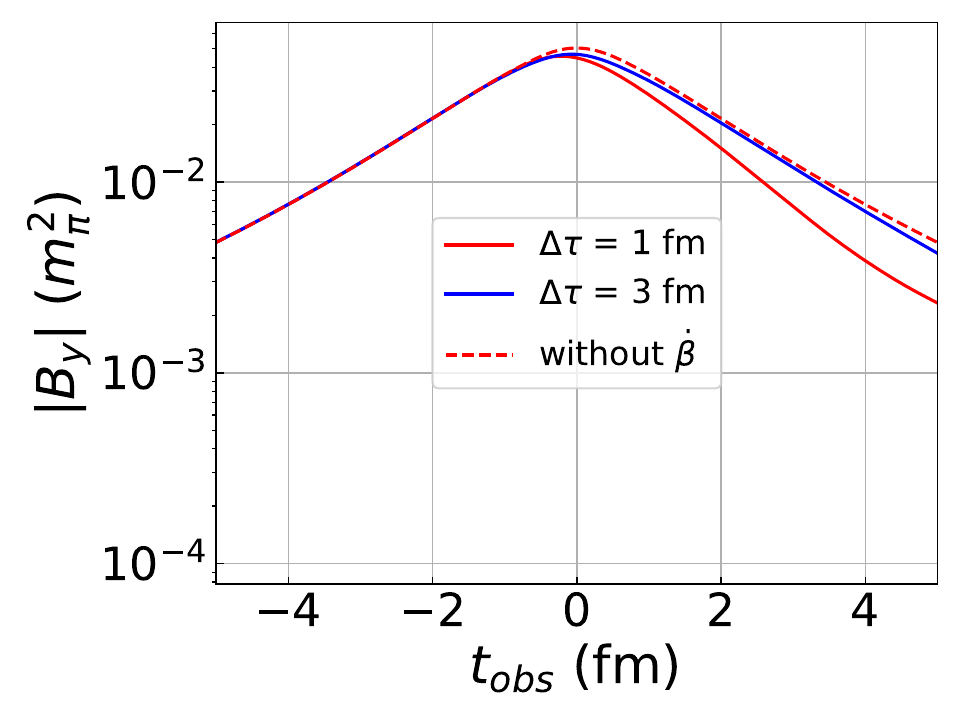}
    \centering
    \includegraphics[width=0.8\linewidth]{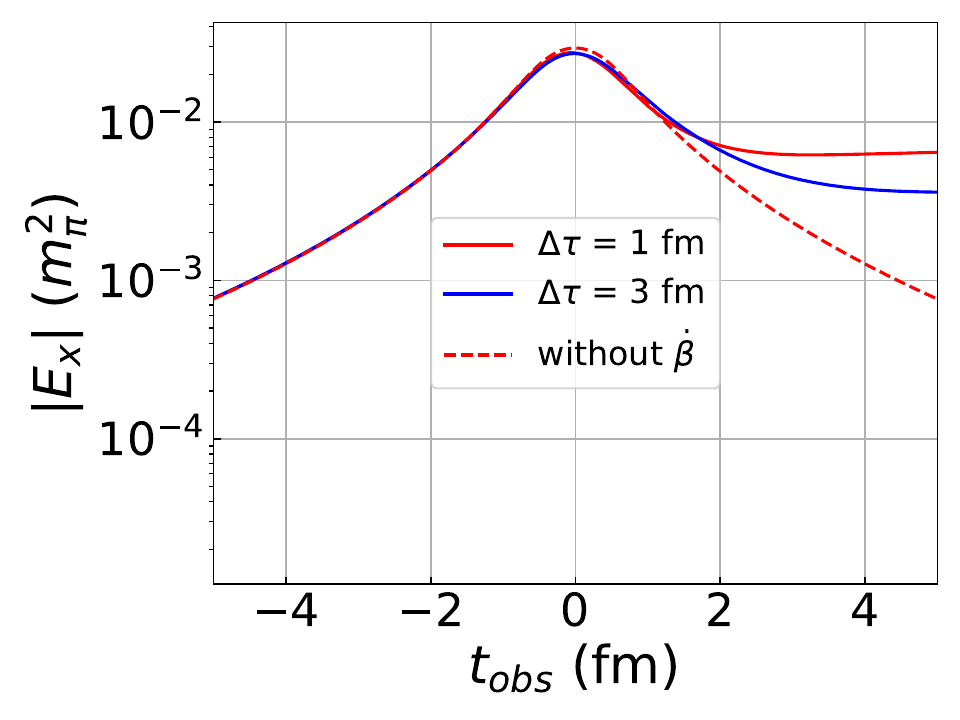}
    \centering
    \includegraphics[width=0.8\linewidth]{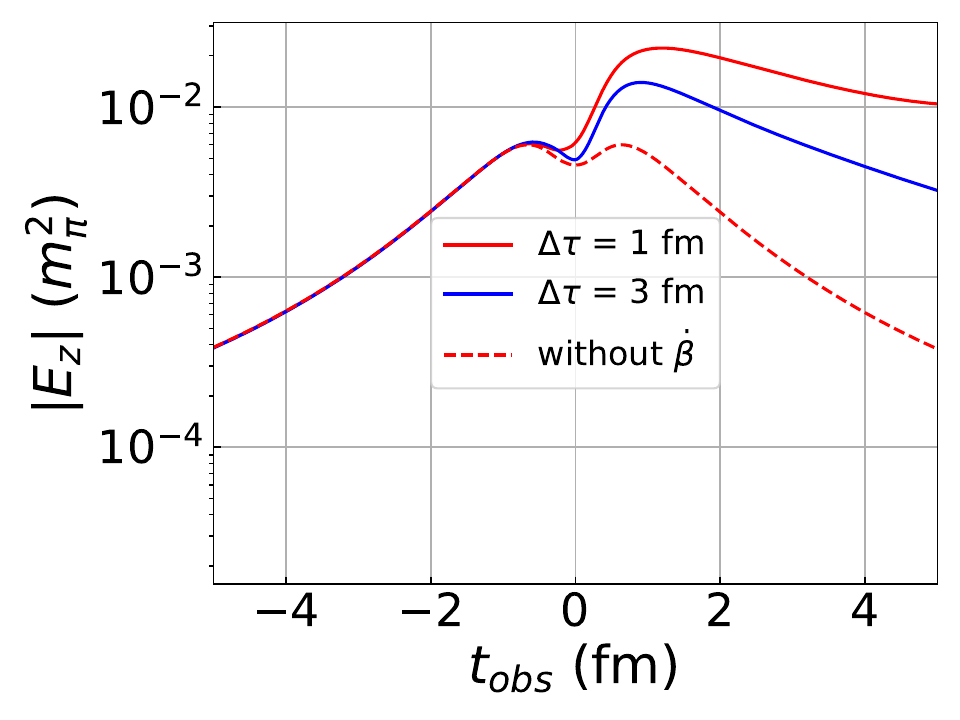}
  \caption{(Color online) Comparison of the components of the fields for $\Delta \tau$ = 1 (solid red lines), 3 fm (solid blue lines) and without $\dot{\beta}$ (red dashed lines) at $\sqrt{s_{NN}}$ = 4 GeV and $b$ = 3 fm at $\textbf{r}_{\text{obs}}$ = (0,0,0). } {\label{fig:comparadtau}}
\end{figure}

We found $|E_z|$ is most sensitive to deceleration and is enhanced substantially after collisions for the baryon stopping scenario compared to the other components of the electric fields. 
In case of no baryon stopping, the peak value of event averaged electromagnetic field rises almost linearly as a function of $\sqrt{s_{NN}}$~\cite{Deng:2012pc}.
To investigate whether this approximate linear dependency on $\sqrt{s_{NN}}$ still holds for baryon stopping scenario, we plot $|B_i|$'s at $({\bf{r}}_{obs}, t_{obs})=0$ as a function of $\sqrt{s_{NN}}$ in Fig.\eqref{fig:rootsvariation}. The top panel is for $b = 3$ fm, and the bottom is for $b = 12$ fm. From the comparison of these two plots, a clear dependence is observed as we go from central to peripheral collisions. However, the approximate linear proportionality with $\sqrt{s_{NN}}$ holds for peripheral collisions; the effect of baryon stopping seems to break the apparent linearity. We found deceleration introduces a small quadratic dependence.
Fig.\eqref{fig:variationwithb} on the other hand illustrates the impact parameter dependence of $|E_i|,|B_i|$'s at $({\textbf{r}}_{\text{obs}}, t_{\text{obs}})=({\bf 0},0)$ for baryon stopping (dashed lines) and without stopping (solid lines). The change in the number of participants with impact parameters causes the observed difference between the two cases.
The parameter $\Delta \tau$ used in Eq.\eqref{eq:partvel} controls the timescale for the deceleration of the participants. We study the effect of varying $\Delta \tau$ on the electromagnetic field in Fig.\eqref{fig:comparadtau}. The results for ($|B_x|$, $|B_y|$) (\text{top two panels}) and ($|E_x|$ and $|E_z|$) (\text{bottom two panels}), are shown for $\Delta \tau = 1$ fm (solid red lines) and $\Delta \tau = 3$ fm (solid blue lines). We kept $b = 3$ fm and $\sqrt{s_{NN}} = 4$ GeV constant in all these cases. We checked that $|E_y|$ is quite similar to that of $|E_x|$ and hence not shown here. These results show a clear dependence of the field strength and its time-evolution on $\Delta \tau$. For comparison, we also show no deceleration results by dashed red lines. 
Naively, one would expect a large $\Delta \tau $ correspond to longer deceleration time results in longer-living fields, but we do not expect any dependency of the peak magnitude of the electromagnetic fields on $\Delta \tau$ for $\tau \leq 0$.
However, in Fig.\eqref{fig:comparadtau}, we found an apparent deviation from this expectation, particularly evident in the bottom panel.
 The peak magnitude of $|E_z|$ appears to be higher for $\Delta \tau = 1$ fm compared to $\Delta \tau = 3$ fm. This disparity primarily arises from the fact that in our parametrization for velocity Eq.\eqref{eq:partvel} a larger $\Delta \tau$ corresponds to a smaller proper deceleration and vice versa and hence varying contribution in the production of EM fields. Moreover, the lumpy charge distribution along the longitudinal direction and the accumulation of these charges around the observation point $\textbf{r}_{\text{obs}}=(0,0,0)$ after collision also depend on $\Delta \tau$. 
The influence of $\Delta \tau$ seems to be more prominent on the electric fields, particularly on $|E_z|$  shown in the bottom panel 
of Fig.\eqref{fig:comparadtau}. 

\begin{figure}[H]
    \centering
    \includegraphics[width=0.8\linewidth]{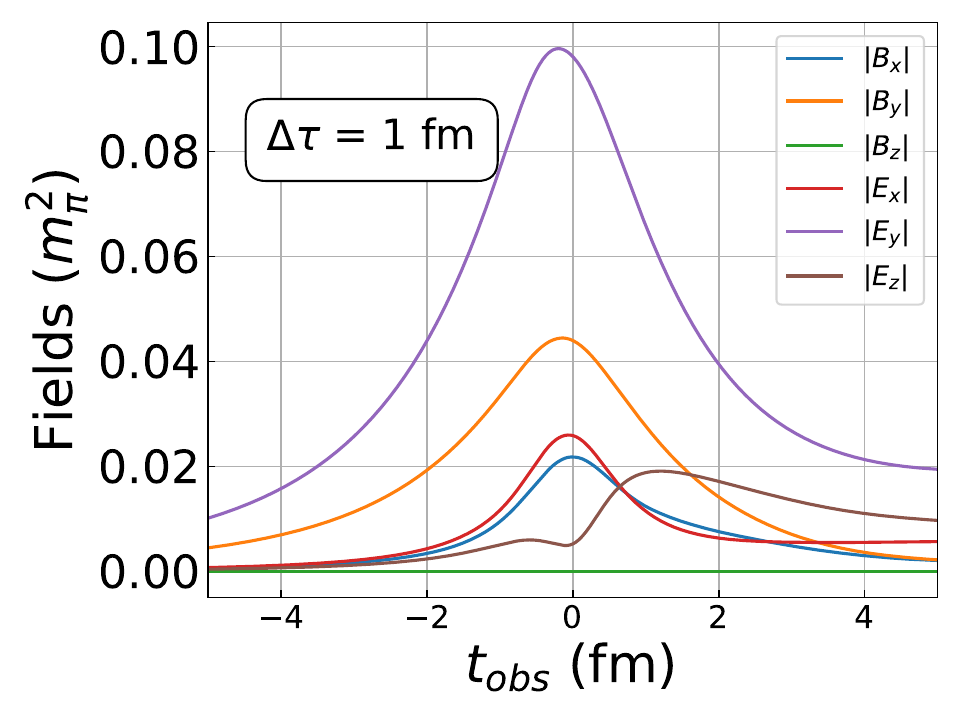}
    \centering
    \includegraphics[width=0.85\linewidth]{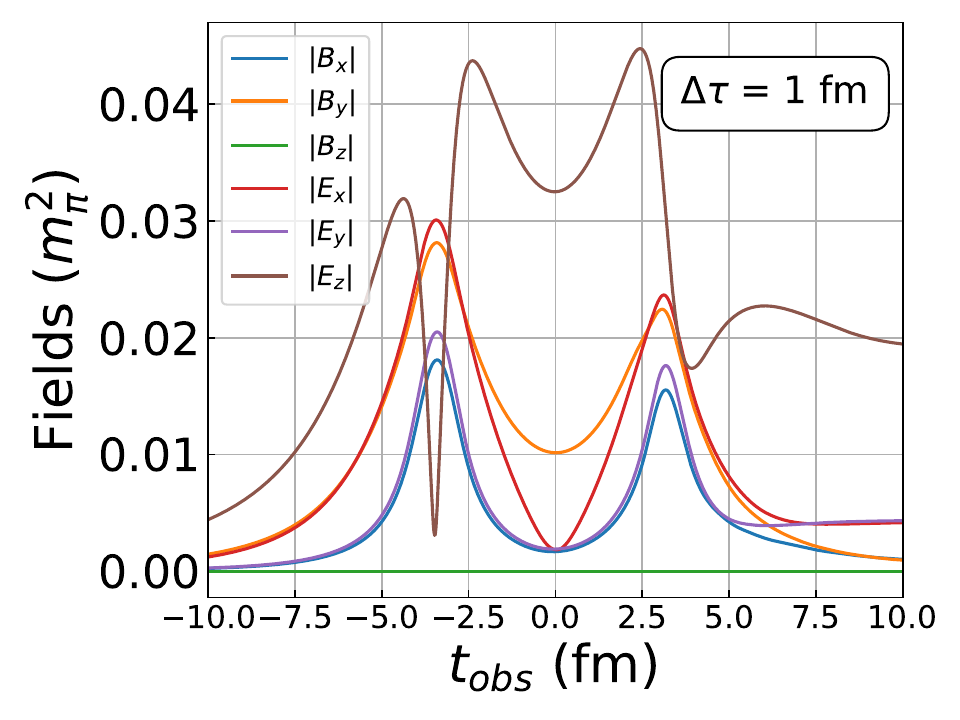}
  \caption{(Color online) Top panel: Fields at ${\bf r}_{\text{obs}}$ = $(0,3,0)$ for $\sqrt{s_{NN}}$= 4 GeV and $\Delta \tau =$ 1 fm.  Bottom panel: same as top panel but for  ${\bf r}_{\text{obs}} = (0,0,3)$.}{\label{fig:diffposition}}
\end{figure}
Up until now, all the results shown were for ${\bf r}_{\text{obs}}$= (0,0,0).  
In Fig.\eqref{fig:diffposition} we show the temporal variation of the fields at two different observation locations ${\bf r}_{\text{obs}}$ = $(0,3,0)$ (top panel) and $(0,0,3)$ (bottom panel) for $\sqrt{s_{NN}}$= 4 GeV and $\Delta \tau$ = 1 fm.
We found the temporal evolution of fields is almost identical for the observation point located on the $y$ axis in the central transverse plane (top panel) as what was observed at ${\bf r}_{\text{obs}}$ = $(0,0,0)$ (Fig.\eqref{fig:temporalevolution}). The only exception is $|E_y|$, which is larger due to the the coherent superposition from the target and projectile.

The passing of target and projectile nuclei through the observation point ${\bf r}_{\text{obs}}=(0,0,3)$ on the $z$ axis is expected to give rise to double-peaked (symmetrically situated around $\tau=0$) structure of the temporal evolution of field components which is apparent from the bottom panel of Fig.\eqref{fig:diffposition}. $|E_z|$ seems to dominate in this case compared to other components. The asymmetry in the field values around $t_{\text{obs}}=0$ fm in Fig.\eqref{fig:diffposition} arises due to the post-collision deceleration of nucleons.

\begin{figure}[H]
    \centering
    \includegraphics[width=0.8\linewidth]{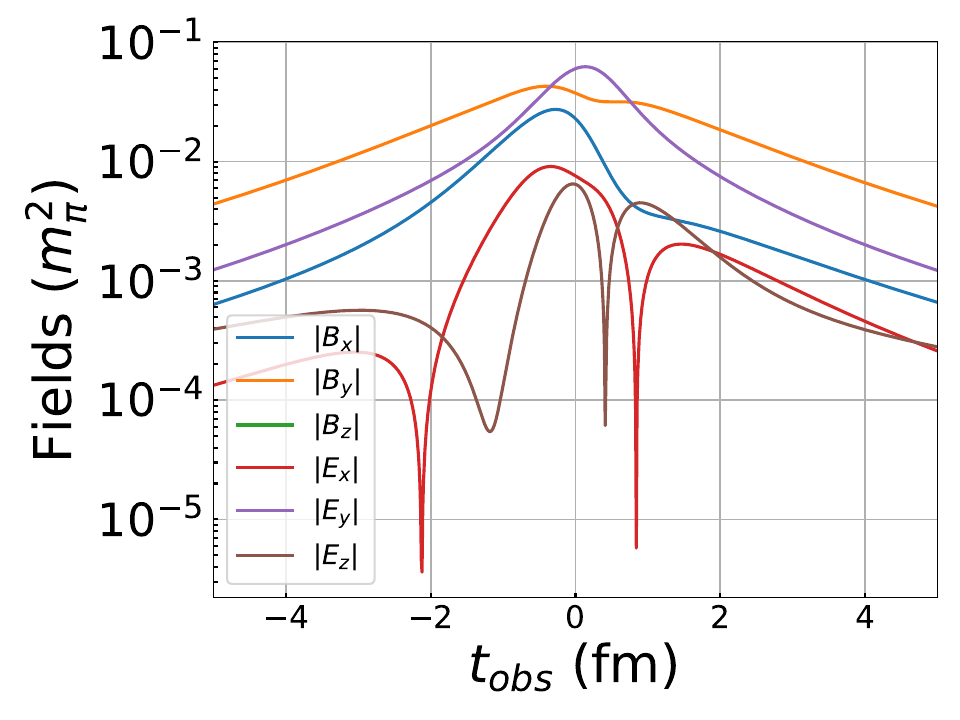}
    \centering
    \includegraphics[width=0.8\linewidth]{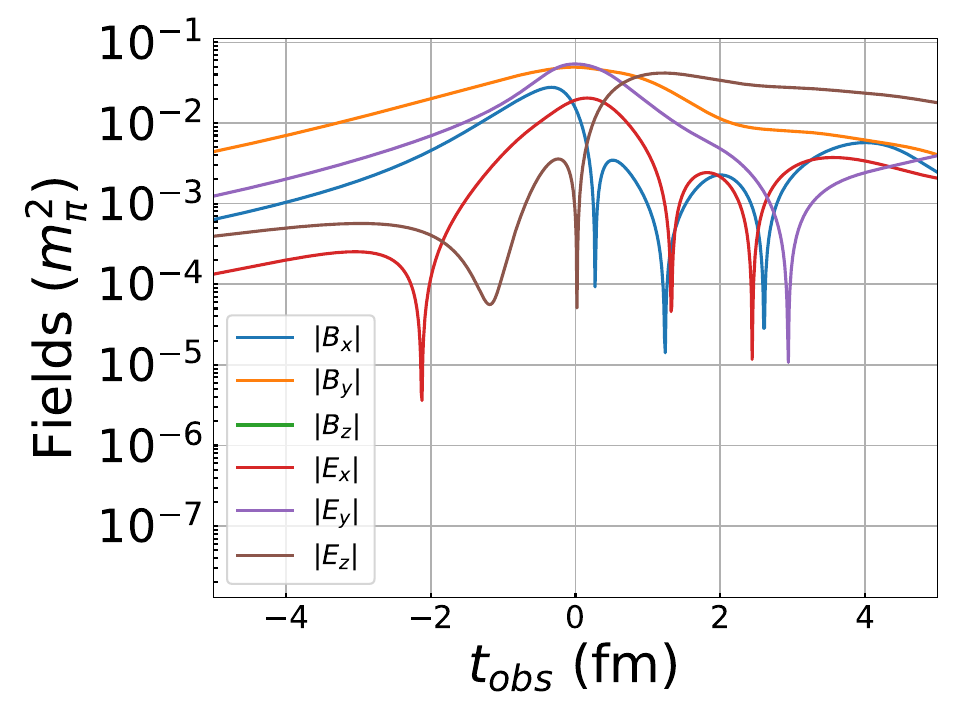}
  \caption{(Color online)  Fields in a randomly selected event. Top panel: without deceleration at ${\bf r}_{\text{obs}}$ = $(0,0,0)$ for $\sqrt{s_{NN}}$= 4 GeV and b= 3 fm. Bottom panel: same as top panel but with deceleration and for the same event as above.}{\label{fig:event1}}
\end{figure}

So far, all the results shown here have been obtained by taking averages of the field components over many (thousand) events. We plotted all six field components for a randomly chosen event for $\sqrt{s_{NN}}$= 4 GeV and b= 3 fm in Fig.\eqref{fig:event1}. The fact that field variations with time is non-trivial due to the lumpy charge distribution and similar magnitude of the components of electric and magnetic fields at ${\bf r}_{\text{obs}}=(0,0,0)$ is apparent from the figure. The top panel corresponds to no deceleration, and the bottom panel corresponds to deceleration case with $\Delta \tau$ = 1 fm. 

\section{Conclusion}\label{conclusion}
In this exploratory study, we use the Monte-Carlo Glauber model with post-collision baryon stopping to investigate the vacuum space-time evolution of electromagnetic fields in low-energy heavy-ion collisions. Several observations are in order: firstly, upon incorporating baryon stopping via a parameterized form
of the velocity of the colliding nucleons, with the key parameter being the time interval for deceleration ($\Delta \tau$), visible effects are observed for $t_{\text{obs}} \ge 0$. 

Secondly, as observed earlier for event-by-event calculations, due to the quantum fluctuations in the nucleon positions, all components of the electromagnetic fields become comparable in high-energy heavy-ion collisions at ($\tau = 0$); in the present study, we found temporal evolution of these electromagnetic fields retain this trend before and after the collision. Specific field components dominate for a given collision impact parameter only when taking the event average. One of the novel finding of the present investigation is that even without any medium effects 
the deceleration enhances electric fields at late times compared to the case of no deceleration; the effect of deceleration is most significant for the longitudinal component of the electric fields at late times.

The presence of deceleration mostly reduces the strength of the magnetic fields post-collisions. However, contrary to the other two components, we found $B_x$ increases at late times compared to the scenario of zero deceleration. 
We observed a slight shift of the peak position of EM fields from $\tau=0$ for the baryon stopping scenario, perhaps akin to the particular form of the parameterized velocity used in this study. 
For higher $\sqrt{s_{NN}}$, when the nucleons move with almost constant velocity after each binary collision, a linear proportionality of the peak value of magnetic fields (at $\tau=0$) with $\sqrt{s_{NN}}$ is observed. However, in the presence of deceleration, this approximate linearity seems to be broken slightly at lower energies. 
Owing to the lower velocities of the colliding nucleons for smaller $\sqrt{s_{NN}}$ the crossing time for the two nuclei becomes longer and 
additionally, the fluctuating nucleon positions along the longitudinal direction give rise to a non-trivial variation of the EM field when measured on an event-by-event basis. 

In future studies, several improvements are conceivable. The assumption that nucleons, following collisions, come to an almost complete halt within a few fm could be relaxed, and a more realistic scenario may entail nucleons attaining a reduced velocity post-collision, with the possibility of further energy loss upon subsequent collisions or maintaining this diminished velocity depending on case-by-case. With this improvement, this model may match with the experimentally measured net baryon density with rapidity, thereby facilitating a more accurate estimation of $\Delta \tau$.
During the conclusion of our research, we encountered the article~\cite{Taya:2024wrm}, where the authors also delve into the realm of low-energy collisions, ranging from $\sqrt{s_{NN}} \sim 3$ to 10 GeV, and investigate the impact of baryon stopping utilizing a hadron cascade model. Our study encompasses the effect of baryon stopping at the initial state within the Glauber model, examining its influence on various components of electromagnetic fields. In contrast, their research focuses on the later stages of heavy-ion collisions, with the effect of stopping considered at the final hadronic stage.

\begin{acknowledgments}
A.P. acknowledges the CSIR-HRDG financial support. H.M. would like to thank Sourendu Gupta and Sandeep Chatterjee for discussions. P.B. acknowledges financial support from DAE project RIN 4001. V.R. acknowledges  financial support from SERB (CRG/2023/001309) , Government of India.
\end{acknowledgments}

\bibliography{ref}

\end{document}